\documentclass[sigconf]{acmart}
\def\BibTeX{{\rm B\kern-.05em{\sc i\kern-.025em b}\kern-.08emT\kern-.1667em\lower.7ex\hbox{E}\kern-.125emX}}
    
\usepackage{nicefrac}
\usepackage{siunitx}
\usepackage{array,framed}
\usepackage{booktabs}
\usepackage{
  color,
  float,
  epsfig,
  wrapfig,
  graphics,
  graphicx,
  subcaption
}
\usepackage{textcomp}
\usepackage{setspace}
\usepackage{latexsym,fancyhdr,url}
\usepackage{enumerate}
\usepackage{algorithm2e}
\usepackage{algpseudocode}
\usepackage{graphics}
\usepackage{xparse} 
\usepackage{xspace}
\usepackage{multirow}
\usepackage{csvsimple}
\usepackage{balance}

\usepackage{
  tikz,
  pgfplots,
  pgfplotstable
}
\usepackage{hyperref}

\usetikzlibrary{
  shapes.geometric,
  arrows,
  external,
  pgfplots.groupplots,
  matrix
}

\pgfplotsset{compat=1.9}

\usepackage{mathtools}

\DeclareMathAlphabet{\mathcal}{OMS}{cmsy}{m}{n}

\DeclareGraphicsExtensions{%
    .png,.PNG,%
    .pdf,.PDF,%
    .jpg,.mps,.jpeg,.jbig2,.jb2,.JPG,.JPEG,.JBIG2,.JB2}

\usepackage{graphicx}
\usepackage{textcomp}
\usepackage{verbatim}
\usepackage{enumitem}
\usepackage{url}
\usepackage{colortbl}
\usepackage{fp}
\usepackage{caption}
\pgfplotsset{width=\linewidth,compat=1.9}

\DeclareMathOperator*{\argmax}{arg\,max}

\newcommand{\vcenteredinclude}[1]{\begingroup
\setbox0=\hbox{\includegraphics[height=3em]{#1}}%
\parbox{\wd0}{\box0}\endgroup}

\begin{document}
\fancyhead{}
\def\thetitle{Segmentation Fault: A Cheap Defense Against Adversarial Machine Learning}
\title{\thetitle}

\author{Doha Al Bared}
\affiliation{%
\institution{American University of Beirut (AUB)}
\city{Beirut}
\country{Lebanon}
}
\email{dma64@mail.aub.edu}

\author{Mohamed Nassar}
\authornote{m.n. started this work at the American University of Beirut (AUB)}
\affiliation{%
\institution{University of New Haven}
\city{West Haven}
\state{CT}
\country{USA}
}
\email{mnassar@newhaven.edu}

\date{}

\begin{abstract}
Recently published attacks against deep neural networks (DNNs) have stressed the importance of methodologies and tools to assess the security risks of using this technology in critical systems. Efficient techniques for detecting adversarial machine learning helps establishing trust and boost the adoption of deep learning in sensitive and security systems. 

In this paper, we propose a new technique for defending deep neural network classifiers, and convolutional ones in particular. Our defense is cheap in the sense that it requires less computation power despite a small cost to pay in terms of detection accuracy. The work refers to a recently published technique called ML-LOO. We replace the costly pixel by pixel leave-one-out approach of ML-LOO by adopting coarse-grained leave-one-out. We evaluate and compare the efficiency of different segmentation algorithms for this task. Our results show that a large gain in efficiency is possible, even though penalized by a marginal decrease in detection accuracy. 
\end{abstract}


\maketitle

\section{Introduction}

Adversarial examples were firstly introduced by Szegedy et al. who demonstrated that by applying a certain hardly perceptible perturbation to an image, one neural network is driven to misclassify the image. 
The reason is that deep neural networks learn input-output mappings that are fairly discontinuous to a significant extent. The attack can be performed by maximizing the network’s prediction error. 
In addition, the specific nature of these perturbations is transferable. The same perturbation can cause the same input to be misclassified by a different network, even when it was trained on a different dataset \cite{szegedy2013intriguing}.

Most methods used to generate adversarial perturbations rely either on gradient-based information or on confidence scores such as class probabilities. For instance, DeepFool \cite{moosavi2016deepfool} relies on an iterative linearization of the classifier to generate minimal perturbations that are sufficient to change classification labels.
However, black-box attacks that are solely based on the final model decision have been recently shown feasible \cite{brendel2017decision}. One strategy consists
in training a local model to substitute for the target network,
using inputs synthetically generated by an adversary and
labeled by the target network \cite{papernot2017practical}. The local substitute  is used to craft adversarial examples, that found to be misclassifed by the targeted deep network.


Adversarial machine learning imposes a threat in real world applications concerning people's safety. For example, an adversarial attack that fools traffic sign classification in autonomous cars may cause devastating consequences. 
It is risky to apply neural networks in security-critical areas today without adequate defense techniques and adversarial training. 
That's why defense and detection of adversarial attacks is currently a research hotspot.
Several defense mechanisms have been proposed to this date such as regularization and data augmentation. A new
variant of distillation was proposed in  \cite{papernot2016distillation} to provide for defense training.
The basic idea is to smooth the network representative function by transferring knowledge encoded into soft probability vectors.  The soft labels are obtained through training with the same network architecture and "Temperature" parameter. 

More similar to our work is Feature Squeezing \cite{xu2017feature}.
Feature squeezing coalesces samples that correspond
to many different feature vectors in the original space into a single sample. This can be achieved for instance by reducing the
color bit depth of each pixel or spatial smoothing. This technique aims at reducing the search space available to an adversary. By comparing the prediction of two DNN models: the original one and the input-squeezed one, 
adversarial examples are shown to be detected with high accuracy and few false positives.
A similar idea appeared in \cite{zhang2018detecting} based on saliency maps and is connected to the concept of explainability in machine learning. 
At visual level, saliency is the regions of the image 
that were particularly influential to the final classification.
The saliency maps of adversarial samples appear to be different than those derived from benign samples. The authors train a binary classifier benign/adversarial based on saliency maps. 

Multi Layer - Leave One Out (ML-LOO) \cite{yang2020ml} proposes another idea using feature (pixel) attribution rather than saliency. Adversarially crafted samples manifest differences in feature attribution histograms when compared to those of original samples. These differences can be captured in form of Inter Quartile Range (IQR) values. A binary classifier can be further trained to flag suspicious input samples.
The approach can work in black box mode where only class probabilities are used, or in white box (Multi-Layer) mode where feature attribution with respect to internal units' values are used as well. The results of ML-LOO are promising based on an extensive set of experiments, but we noticed a significant computational cost. 

In this paper, we revisit ML-LOO and propose to support more coarse-grained features by using classical image segmentation techniques. We evaluate the segmentation techniques and compare to the original work in \cite{yang2020ml}. 

The rest of the paper is organized as follows: in Section \ref{related} related work is surveyed. We revisit ML-LOO and present our implementation in Section \ref{revisit}. We present our "Segmentation Fault" approach in Section \ref{segfault}. Finally Section \ref{concl} concludes the paper and outlines future work. 

\section{Related Work} 
\label{related}

Neural networks, deep convolutional ones in particular, are currently the best machine learning algorithms at object classification. However, these models have been shown to be vulnerable to many adversarial deceptions. 
By slightly changing the pixel values of an image, one can lead a convolutional neural network to make wrong predictions \cite{szegedy2013intriguing}. Attacks can be divided into model-specific (white box) and model-agnostic (black box). Model-specific assumes access to the gradients of the learning model \cite{kurakin2016adversarial}. Model-agnostic techniques transact with classifiers as black boxes and do not require internal knowledge of the model or the training data \cite{papernot2017practical}.
Some adversarial techniques lead to stealthy, unperceived to the human observer, samples. For example, changing only one pixel in the input image is shown to be enough in \cite{su2019one}. However, the feature attribution of this pixel will be significant. Recently, more interest is put on physical samples such as adversarial patches and adversarial 3D printed artifacts \cite{athalye2018synthesizing}. 
We showed how adversarial patches may help shoplifting smart stores in \cite{nassar2019shoplifting}. 
A recent review of adversarial attacks and defenses in the physical world is in \cite{renadversarial}. A nice summary of the history of adversarial machine learning is in \cite{biggio2018wild}. ML-LOO \cite{yang2020ml}, a recently proposed defense technique, shows applicability over a wide set of attacks and datasets. We carry on the same direction of ML-LOO and propose image segmentation to boost the performance of this approach.

\section{ML-LOO: Background \& Implementation}
\label{revisit}
The idea of Leave One Out (LOO) \cite{yang2020ml} is simply to take an $m \times n$ pixels image and sample $m \times n$ images out of it by blacking out pixels one by one. Sometimes, $m \times n \times 3$ are sampled by blacking R, G and B channels of each pixel. The sample images are forward-propagated through the neural network classifier in question. The output values of of a predefined set of network nodes are recorded. Several operational modes are available: 
\begin{description}
    \item[1D] Only the probability of the predicted class for the original image is recorded: 
    $f(x_i)_c$ where $x$ is the original image, $x_i$ is the image with  blacked pixel $i$, $f$ is the neural network representative function, and $c= \argmax_{j\in C} f(x)_j$. $C$ is the set of output classes. 
    \item[Output layer] Only the output layer values are recorded. 
    \item[Multi Layer (ML-LOO)] The values for the output layer and a randomly pre-selected set of hidden nodes from different intermediate layers are recorded.
\end{description}
The $m \times n$ recorded values for a node are collected. More precisely we record the absolute difference of the LOO values and the original single value $\phi(x)_{i,\text{node}_j} = \vert f(x_i)_{\text{node}_j} -  f(x)_{\text{node}_j} \rvert$. The original value corresponds to the value of the node taken from the forward propagation of the original image. This difference is called \textit{feature attribution}.
We abstract the group of these recorded differences for all pixels $i$ and a single node $j$ in one statistical measure, namely the Interquartile range (IQR).
$$ IQR (\phi(x)_{\text{node}_j}) = Q_{\phi(x)_{\text{node}_j}}(0.75) - Q_{\phi(x)_{\text{node}_j}}(0.25) $$
where: $$ Q_{\phi(x)_{\text{node}_j}}(p) = \min \{\beta:\frac{\#\{i :\phi(x)_{i,\text{node}_j}< \beta\}}{m \times n} \geq p \} $$
The IQR is useful for measuring the dispersion of feature attributions. The authors of \cite{yang2020ml} observed that adversarial samples cause a larger dispersion. The intuition is that in normal images a few "foreground" pixels have larger feature attributions compared to a majority "background" pixels. These pixels with high feature attribution serve as an explanation of the classifier decision. The feature attributions are more dispersed in adversarial images, which can be captured by a wider IQR value.  We are a bit more cautious in generalizing this observation as a common effect for all adversarial techniques. 
Still, we postulate that in multi-dimensional IQR, interquartile values computed over neurons across the output layer and many hidden layers in the network, may serve as an interesting  feature vector representation of an image.  The IQR feature vectors for a large and varied labeled dataset of normal and adversarial samples can be used to train a simple machine learning binary classifier. 
The binary classifier task is to take an input image, as represented by its IQR feature vector, and decides whether it is benign or malicious. Experiments show that ML-LOO efficiently filters the majority of adversarial samples.

We however notice that ML-LOO is not cheap in terms of computational performance as it requires $m \times n$ forward-propagations per image. We discuss our implementation of this proposed  technique and findings in this section and differ our approach to boost its computational speed to the next section. 

We experimented with CIFAR10 dataset \cite{krizhevsky2009learning} consisting of 32x32 tiny images  (e.g. 
\vcenteredinclude{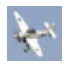}, \vcenteredinclude{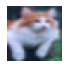},
\vcenteredinclude{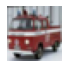}), 10 classes (airplane, automobile, bird, cat, deer, dog, frog, horse, ship, truck), and 60,000 examples 
separated into training (50,000) and testing (10,0000) subsets. Our CIFAR10 pretrained classifier is a deep convolutional neural network with 26 layers and 309,290 parameters\footnote{Skeleton code from \url{https://www.kaggle.com/c/cifar-10/discussion/40237}}. The model has 94\% training accuracy and 89\% testing accuracy.  We used Foolbox \cite{rauber2017foolboxnative,rauber2017foolbox} to generate adversarial samples. Four attack algorithms have been evaluated: Fast Gradient Sign Method (FGSM) \cite{goodfellow2014explaining}, PGD \cite{madry2017towards}, DeepFool \cite{moosavi2016deepfool},  and C\&W \cite{carlini2017towards}. 
The efficiency of these attacks for different parameters is shown in Table \ref{tab:adv}. For C\& W we left epsilons as None and shorten the number of steps to 500 for lessening the generation time.

\begin{table}[tbp]
\centering
\caption{Neural Networks Model Classification Accuracy for One Random Batch of 128 Adversarial Images \label{tab:adv}}
\begin{tabular}{|l|c|c|c|}
\hline
 Attack $\downarrow \backslash$ $\epsilon \xrightarrow{}$ & $\epsilon=0.02$ & $\epsilon=0.06$ & $\epsilon=0.1$\\
  \hline   
    FGSM  & 41.41\%  & 13.28\% & 11.72\% \\
    PGD & 32.03\% & 1.56\% &  0.0\%\\
    DeepFool & 40.62\% & 11.72\%&   10.94\%\\
    \hline
    C\&W  &  \multicolumn{3}{|c|}{17.19\%}\\
    \hline 
    No attack &  \multicolumn{3}{|c|}{95.31\%} \\
    \hline
\end{tabular}
\end{table}

We implemented 3 flavors of ML-LOO:
\begin{description}
\item [1D] Only taking the IQR of the predicted-class output node, 
\item [10D]Taking the IQR vector for the 10 nodes of the output layer, 
\item [4K-D] The IQR vector has 4K values: 200 nodes are randomly chosen from each layer in the set of the last 20 layers of the network (since the last layer is only 10 nodes, the real number is 3810 values). 
\end{description}

\begin{figure}[t]
    \centering
    \includegraphics[width=\linewidth]{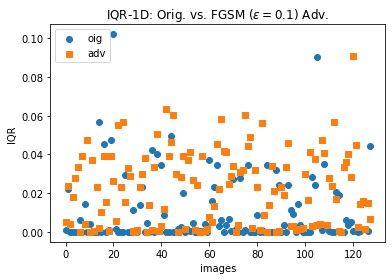}
    \caption{IQR-1D: Adversarial images show larger IQR values in average}
    \label{fig:IQR1D}
\end{figure}

\begin{figure*}[t]
    \centering
    \includegraphics[width=\textwidth]{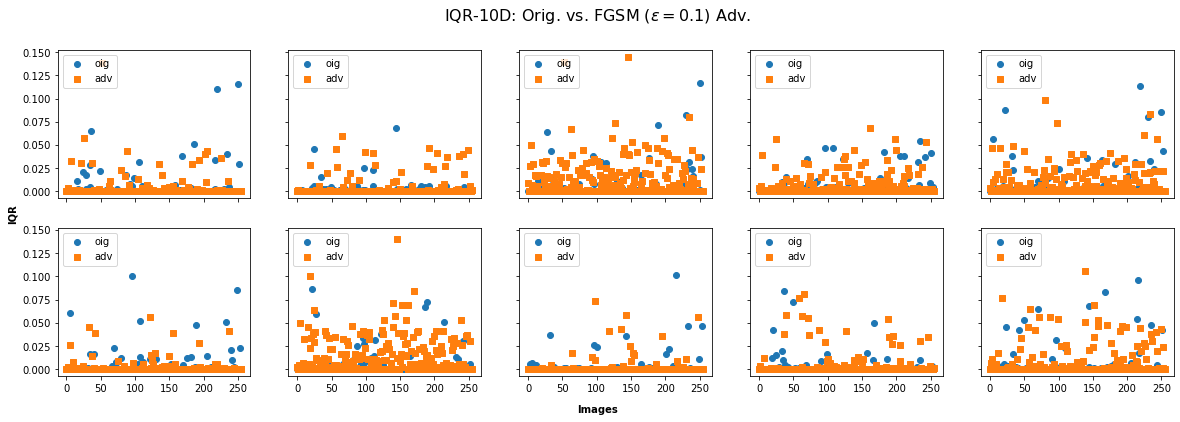}
    \caption{IQR-10D: Adversarial images show larger IQR values in average for most of the dimensions}
    \label{fig:IQR10D}
\end{figure*}

A scatter plot for a batch of 128 original images and 128 adversarial images obtained by FGSM ($\epsilon=0.1$) is shown in Figure \ref{fig:IQR1D}. The attack was successful in reducing the classifier accuracy to around 10\% for the adversarial samples, making it no better than a random classifier with 10 classes. Scatter plots for the dimensions in IQR 10D are show in Figure \ref{fig:IQR10D}. 

Furthermore, our results for the three aforementioned IQR models, adversarial techniques and off-the-shelf binary classifiers are presented in Table \ref{tab:MLLOO_Res}. For each attack/experiment, we use 30 batches of benign images and their counterpart 30 batches of adversarial samples ($30*128=3840$ images) generated using any misclassification as criteria and multiple epsilons. Each third of adversarials (10 batches) is using a different epsilon: 0.02, 0.06, or 0.1. For C\&W we keep epsilons free to choose by the algorithm and limit the number of steps to 500. 80\% of the data were used for training the classifier and only 20\% for testing. The values shown correspond to the classifiers performance over the testing data. 

\newcommand{\fcell}[1]{%
\FPeval{\grayshade}{(1.5-#1)*1.1}%
\xdef\tempa{\grayshade}%
\cellcolor[gray]{\tempa} #1} 

\begin{table*}[t]
    \centering
    \caption{Off-the-shelf classifiers detection AUC for the different IQR representations}
    \begin{tabular}{|c|c|c|c||c|c|c||c|c|c||c|c|c|}
    \cline{2-13}
          \multicolumn{1}{c|}{}& \multicolumn{12}{|c|}{Attacks}  \\
         \cline{2-13} 
         \cline{2-13}
          \multicolumn{1}{c|}{}& \multicolumn{3}{|c||}{\textbf{FGSM}} & \multicolumn{3}{|c||}{\textbf{PGD}} & \multicolumn{3}{|c||}{\textbf{DeepFool}} & \multicolumn{3}{|c|}{\textbf{C\&W}} \\
         \hline 
         Classifier & \emph{1-D} & \emph{10-D} & \emph{4K-D} & \emph{1-D} & \emph{10-D} & \emph{4K-D} & \emph{1-D} & \emph{10-D} & \emph{4K-D} &\emph{1-D} & \emph{10-D} & \emph{4K-D} \\
         \hline
         \textbf{Random Forest} & \fcell{0.633} & \fcell{0.765} & \fcell{0.854} & \fcell{0.693} & \fcell{0.782} & \fcell{0.685} & \fcell{0.671} & \fcell{0.792} & \fcell{0.649} & \fcell{0.917} &  \fcell{0.929} & \fcell{0.660} \\ 
         \textbf{SVC} & \fcell{0.738} & \fcell{0.768} & \fcell{0.802} & \fcell{0.720} & \fcell{0.735}& \fcell{0.744} & \fcell{0.745} &\fcell{0.755} & \fcell{0.746}& \fcell{0.926} & \fcell{0.930 }& \fcell{0.819}\\
         \textbf{XGB} & \fcell{0.737}  & \fcell{0.810}& \fcell{0.838} & \fcell{0.740} & \fcell{0.785} & \fcell{0.786} & \fcell{0.750} &\fcell{0.811}  & \fcell{0.836}& \fcell{0.946} & \fcell{0.956} &  \fcell{0.956}\\
         \hline 
    \end{tabular}
    \label{tab:MLLOO_Res}
\end{table*}

The best detection accuracies are recorded against the C\&W attack. 
Results shows that increasing the IQR representation dimensionality improves the detection accuracy consistently when XGB is used, even though the random forest classifier and sometimes SVC badly handle the case of high dimensional 4K features. 
The shown numbers belong to the default parameters of the classifiers. However, we noticed that these numbers can be significantly enhanced by a grid search for the best parameters, especially for the high-dimensional (4K) case. For example, a grid search for the best $C$ and $\gamma$ parameters of the SVC classifier showed an improvement of 3\% in detection accuracy for the FGSM attack and 4K-D features. 

\section{Segmentation Fault}
\label{segfault}
It is worth noting that the time taken to compute the feature attribution values for a batch of 128 images is in the order of one minute for IQR-1D and IQR 4k-D. The time is a bit less for IQR-10D since no post-processing is required. This time is independent of whether the batch is benign or adversarial. We used a Tesla P100-PCIE-16GB GPU from Google Colab. At each iteration we black-out one pixel for all the images in the batch and forward propagate the batch at once. Also note the considerable memory requirements for IQR-4K-D which is in the order of \verb|BATCH_SIZE*W*H*4000|. For CIFAR10 we need around 2GB to store the feature attribution vectors for a batch of 128 images, before reducing it to IQR vectors of size 4000 each (\verb|BATCH_SIZE*4000|). 
Our approach consists on improving upon these processing and memory requirements by using segmentation. The idea is to simply to black out one segment at a time, reducing the number of blackouts from $w \times h$ to \verb|nb_segments|, hence reducing the processing and memory requirements by a ratio of 10 to 100 in the case of CIFRA10 tiny images, and even more in the case of a larger image dataset. However, it is not clear which segmentation technique would be more successful on preserving a good detection accuracy, hence the motivation of our work.

We consider the following segmentation methods, and used their scikit-image implementation \cite{scikit-image}. 

\begin{table}[htbp]
\centering
\caption{Examples of Felzenszwalb’s  Segmentation\label{tab:Felzen}}
\begin{tabular}{c|cccc}
  \hline 
\textbf{Scale}   & 1 & 10 & 100 & 1000 \\
  \hline
  \textbf{Label} & \multicolumn{4}{c}{Truck} \\
  \textbf{\#Segments}  & 15 & 15 & 16 & 7 \\ 
  \textbf{Image} & \vcenteredinclude{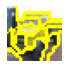} & \vcenteredinclude{pics/truck1.png} & \vcenteredinclude{pics/truck1.png}& \vcenteredinclude{pics/truck1.png} \\
  \hline 
    \textbf{Label} & \multicolumn{4}{c}{Automobile} \\
  \textbf{\#Segments}  & 18 & 18 & 19 & 12 \\ 
  \textbf{Image} & \vcenteredinclude{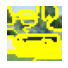} & \vcenteredinclude{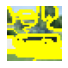} & \vcenteredinclude{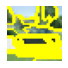}& \vcenteredinclude{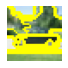} \\
  \hline
  \textbf{Label} & \multicolumn{4}{c}{Deer} \\
  \textbf{\#Segments}  & 11 & 11 & 11 & 5 \\ 
  \textbf{Image} & \vcenteredinclude{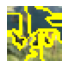} & \vcenteredinclude{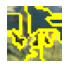} & \vcenteredinclude{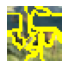}& \vcenteredinclude{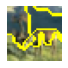} \\
  \hline
\end{tabular}
\end{table}
\begin{table}[htbp]
\centering
\caption{ Examples of QuickShift  Segmentation\label{tab:quickshift}}
\begin{tabular}{|l|cc|cc|cc|}
\multicolumn{7}{c}{\textbf{\#Segments} } \\
\hline
 labels $\xrightarrow{}$ & \multicolumn{2}{c|}{Airplane} & \multicolumn{2}{c|}{Bird} & \multicolumn{2}{c|}{Ship}\\
 \hline 
$ \sigma \downarrow \backslash m \xrightarrow{}$     & 10 & 20  & 10 & 20  & 10 & 20 \\
  \hline   
    0  & 61& 12& 224& 24 & 170 & 25\\
    1 & 35& 12& 87& 19 & 56& 11 \\
    2 & 16& 6& 30 & 11 & 22& 6 \\
    \hline
\end{tabular}
\end{table}
\begin{table}[htbp]
\centering
\caption{Examples of SLIC Segmentation\label{tab:slic}}
\begin{tabular}{c|cccc}
  \hline 
\textbf{n\_segments}  & 1 & 32 & 64 & 128  \\
  \hline
  \textbf{Label} & \multicolumn{4}{c}{Ship} \\
  \textbf{Actual \#Segments}  & 1 & 18 & 48 & 98 \\ 
  \textbf{Image} & \vcenteredinclude{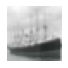} & \vcenteredinclude{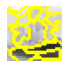} & \vcenteredinclude{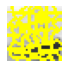}& \vcenteredinclude{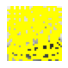} \\
  \hline 
    \textbf{Label} & \multicolumn{4}{c}{Automobile} \\
  \textbf{Actual \#Segments}  & 1 & 13 & 33 & 79 \\ 
  \textbf{Image} & \vcenteredinclude{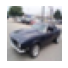} & \vcenteredinclude{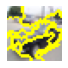} & \vcenteredinclude{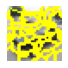}& \vcenteredinclude{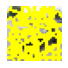} \\
  \hline
  \textbf{Label} & \multicolumn{4}{c}{Frog} \\
  \textbf{Actual \#Segments}  & 1 & 19 & 38 & 91 \\ 
  \textbf{Image} & \vcenteredinclude{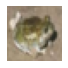} & \vcenteredinclude{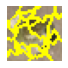} & \vcenteredinclude{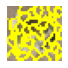}& \vcenteredinclude{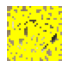} \\
  \hline
\end{tabular}
\end{table}
\begin{description}
\item [Felzenszwalb’s Segmentation \cite{felzenszwalb2004efficient}] This algorithm has a single parameter called \textit{scale} (default value = 1). Higher scale means
less and larger segments. The actual segment size and number of segments can vary depending on local contrast. Some examples for random images from CIFAR10 are shown in Table \ref{tab:Felzen}.

\item[Quickshift Segmentation \cite{vedaldi2008quick}] 
This algorithm computes a hierarchical segmentation over five dimensions (i.e. 3D color, width and height). It has two main parameters: \textit{sigma} (default value = 0) controls the scale of the local density approximation, and \textit{max\_dist} (default value = 10) selects a level in the hierarchical segmentation that is produced. Higher max\_dist and sigma means less segments. Some examples for random images from CIFAR10 and different values of max\_dist ($m$) and sigma ($\sigma$) are shown in Table \ref{tab:quickshift}.

\item[SLIC \cite{achanta2012slic}] 
Simple Linear Iterative Clustering (SLIC) performs K-means clustering in the 5d space of color information (3 dimensions) and image location (x,y). The main parameter is n\_segments which corresponds to the number of centroids for K-means. Some examples for random images from CIFAR10 and different values of n\_segments and  are shown in Table \ref{tab:slic}. 
\end{description}

\begin{table*}[t]
    \centering
    \caption{Off-the-shelf XGBoost classifier detection AUC for different IQR representations and different segmentation techniques}
    \begin{tabular}{|c|l|c|c|c||c|c|c||c|c|c||c|c|c|}
    \cline{3-14}
          \multicolumn{2}{c|}{}& \multicolumn{12}{|c|}{Attacks -- epsilons $=[0.02, 0.06, 0.1]$}  \\
          
         \cline{3-14} 
         \cline{3-14}
          \multicolumn{2}{c|}{}& \multicolumn{3}{|c||}{\textbf{FGSM}} & \multicolumn{3}{|c||}{\textbf{PGD}} & \multicolumn{3}{|c||}{\textbf{DeepFool}} & \multicolumn{3}{|c|}{\textbf{C\&W}} \\
         \hline 
         \multicolumn{2}{|c|}{Segmentation $\downarrow$}  & \emph{1-D} & \emph{10-D} & \emph{4K-D} & \emph{1-D} & \emph{10-D} & \emph{4K-D} & \emph{1-D} & \emph{10-D} & \emph{4K-D} &\emph{1-D} & \emph{10-D} & \emph{4K-D} \\
         \hline
         \multirow{4}{*}{\textbf{Felzenszwalb}} & scale = 1 &  \fcell{0.707} & \fcell{0.782} & \fcell{0.851} & \fcell{0.688} & \fcell{0.749} & \fcell{0.804} & \fcell{0.746} & \fcell{0.790}  & \fcell{0.822} & \fcell{0.877}&\fcell{0.936}  &\fcell{0.930} \\
        & scale = 10 & \fcell{0.705} & \fcell{0.797} & \fcell{0.846} & \fcell{0.692}  & \fcell{0.753} & \fcell{0.766} & \fcell{0.762}  & \fcell{0.816} & \fcell{0.838} &   \fcell{0.887} & \fcell{0.927}&\fcell{0.916}  \\
          & scale = 100 &\fcell{0.695}  & \fcell{0.814} &\fcell{0.834}  &  \fcell{0.688} & \fcell{0.754} & \fcell{0.795} & \fcell{0.762}  & \fcell{0.811}  & \fcell{0.807} & \fcell{0.887}& \fcell{0.926} & \fcell{0.925} \\
          & scale = 1000 & \fcell{0.632} & \fcell{0.782} & \fcell{0.830} & \fcell{0.593} & \fcell{0.665} & \fcell{0.763} & \fcell{0.672} &\fcell{0.735}  & \fcell{0.775} &\fcell{0.797} & \fcell{0.825}& \fcell{0.825}\\
         \hline 
         \hline
         \multirow{6}{*}{\textbf{QuickShift}} & $\sigma=0, m=10$  & \fcell{0.726} & \fcell{0.817} & \fcell{0.853} & \fcell{0.744} & \fcell{0.781} & \fcell{0.771} & \fcell{0.776} & \fcell{0.839} & \fcell{0.835} &  \fcell{0.944}& \fcell{0.946} & \fcell{0.940}\\
          & $\sigma=0, m=20$  & \fcell{0.694} & \fcell{0.797} &  \fcell{0.842} & \fcell{0.703} & \fcell{0.782} & \fcell{0.782} & \fcell{0.761} & \fcell{0.808} &  \fcell{0.827} &  \fcell{0.923}& \fcell{0.937} & \fcell{0.932} \\
          & $\sigma=1, m=10$ & \fcell{0.738} & \fcell{0.798} & \fcell{0.829} & \fcell{0.711} & \fcell{0.776} & \fcell{0.765} & \fcell{0.770} & \fcell{0.810} & \fcell{0.801} & \fcell{0.897}& \fcell{0.923}&  \fcell{0.923}\\
          & $\sigma=1, m=20$ & \fcell{0.710} & \fcell{0.816} & \fcell{0.849} & \fcell{0.637} & \fcell{0.759} & \fcell{0.767} & \fcell{0.753} & \fcell{0.805} & \fcell{0.792} &\fcell{0.877} & \fcell{0.941}& \fcell{0.925}\\
         & $\sigma=2, m=10$ &  \fcell{0.720} & \fcell{0.808} & \fcell{0.849} & \fcell{0.664} & \fcell{0.736}   & \fcell{0.757} &\fcell{0.770} & \fcell{0.806}  & \fcell{0.834} & \fcell{0.895} & \fcell{0.922} & \fcell{0.925} \\
         & $\sigma=2, m=20$ & \fcell{0.692} & \fcell{0.799}& \fcell{0.839} & \fcell{0.615} & \fcell{0.694} & \fcell{0.744}  & \fcell{0.706} & \fcell{0.774} & \fcell{0.790} & \fcell{0.833}& \fcell{0.897}& \fcell{0.879} \\
         \hline
         \hline
          \multirow{4}{*}{\textbf{SLIC}} & n\_segments = 32 &  \fcell{0.727} & \fcell{0.814} & \fcell{0.864} &\fcell{0.682}  & \fcell{0.755} & \fcell{0.755} &  \fcell{0.774} &\fcell{0.805}   & \fcell{0.805} &\fcell{0.895} &\fcell{0.934} & \fcell{0.923}\\
        & n\_segments = 64 & \fcell{0.730} & \fcell{0.809} & \fcell{0.865}  & \fcell{0.723} & \fcell{0.758} & \fcell{0.794} & \fcell{0.782} & \fcell{0.826} & \fcell{0.820} &  \fcell{0.929} & \fcell{0.947}& \fcell{0.941} \\
          & n\_segments = 128 &  \fcell{0.761}&  \fcell{0.812}& \fcell{0.855} & \fcell{0.733} & \fcell{0.773} & \fcell{0.777} & \fcell{0.773} & \fcell{0.826} & \fcell{0.831} & \fcell{0.945}& \fcell{0.959}& \fcell{0.946}\\
         \hline 
         \hline
    \multicolumn{2}{|c|}{\textbf{ML-LOO}} & \fcell{0.728} & \fcell{0.829}& \fcell{0.871}& \fcell{0.748}& \fcell{0.793}& \fcell{0.806}& \fcell{0.793}&\fcell{0.829} & \fcell{0.763}& \fcell{0.918}& \fcell{0.967} & \fcell{0.984} \\
    \hline 
    \end{tabular}
    \label{tab:Seg_Res}
\end{table*}


The results of experimentation with different IQR representations, different segmentation techniques and different adversarial techniques are shown in Table \ref{tab:Seg_Res}. For each attack/experiment, we use 30 batches of benign images and 30 batches of adversarial samples  ($30*128=3840$ images) generated using any misclassification as criteria and multiple epsilons (0.02, 0.06, and 0.1). 80\% of the data were used for training the classifier and only 20\% for testing. The XGB classifier is used off-the-shelf for these experiments. The values shown correspond to the XGB performance over the testing data. 
It can be observed that generally 4K-D improves over 10-D and 10-D improves over 1-D. In general, the accuracy values found are very similar to what can be achieved with ML-LOO, even though slightly lower. The more segments are generated, the more accuracy is obtained, yet SLIC segmentation method looks superior in the sense that it can obtain better accuracy given that equal average number of segments is generated.

The time and memory comparison of ML-LOO versus our approach is shown in Table \ref{tab:comparison}. The time value corresponds to the time for extracting the feature attributions for one batch of 128 images. The size memory value is the size of the object for storing all the feature attributions for one batch of 128 images before being reduced to IQR values. The order of magnitude of the values are represented in grayscale. For time, darker cells correspond to longer processing times. For memory, darker cells correspond to larger memory space required. Results show that for most parameters, our segmentation approach requires much less memory and runs much faster than ML-LOO. We notice that 10-D is consistently the fastest to execute since it does not require any post-processing beyond forward-propagation through the classifier. 1-D has an additional step for fetching the value corresponding to the predicted class for the original image. 4K-D has to loop through the layers to fetch the values of the selected nodes. In terms of memory, 10-D requires 10 times more memory than 1-D and 4K-D requires 381 times more memory than 10-D.

\begin{figure}[t]
\centering
\begin{tikzpicture}
\begin{axis}[
    legend pos=north west,
    ylabel=Time(s),
    xlabel=AUC,
	enlargelimits=0.05
]
\addplot[    
    scatter/classes={Felzenszwalb={blue}, QuickShift={green}, SLIC={black}, MLLOO={red}},
    only marks,
    scatter,
    mark=halfcircle*,
    mark size=4pt,
    scatter src=explicit symbolic,
    ]
table[meta=Class]{scattered.dat};
\legend{Felzenszwalb, QuickShift, SLIC, ML-LOO};
\end{axis}
\end{tikzpicture}
\caption{Trade-off Time vs. Detection Accuracy \label{scatter1}}
\end{figure}
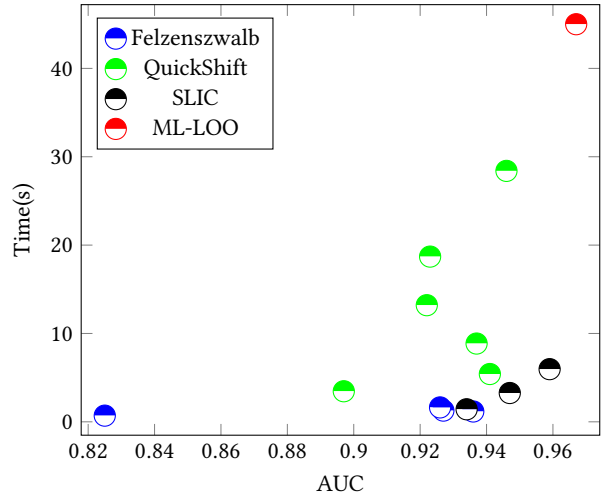

Figure \ref{scatter1} shows an example of the performance of different segmentation techniques and their AUC scores. The scatter plot corresponds to the experiments with the C\&W attack and the 10-D representation. Each point represents a segmentation technique and particular parameters.
Results show that SLIC segmentation is generally better in reducing processing time while scoring AUC scores that are practically the same as those of ML-LOO.

Our ML-LOO implementation and the source code for all the experiments in this paper are available in Google Colab notebooks format at \url{https://github.com/mnassar/segfault}.

\begin{table*}[t]
\newcommand{\fcellsec}[1]{%
\FPeval{\grayshade}{(100-#1)/100}%
\xdef\tempa{\grayshade}%
\cellcolor[gray]{\tempa} #1s} 

\centering
\caption{Comparison of time and memory requirements for representing one batch of 128 images \label{tab:comparison}}
       \begin{tabular}{|c|l|c|c|c||c|c|c|}

         \cline{3-8} 
         \cline{3-8}
          \multicolumn{2}{c|}{}& \multicolumn{3}{|c||}{\textbf{Time}} &  \multicolumn{3}{|c|}{\textbf{Memory}}\\
         \hline 
         \multicolumn{2}{|c|}{Segmentation $\downarrow$}  & \emph{1-D} & \emph{10-D} & \emph{4K-D}  & \emph{1-D} & \emph{10-D} & \emph{4K-D}\\
         \hline
         \multirow{4}{*}{\textbf{Felzenszwalb}} & scale = 1 & \fcellsec{1.79} & \fcellsec{1.17} & \fcellsec{1.77}  & \cellcolor[gray]{0.9} 9.8KB & \cellcolor[gray]{0.9} 98KB & \cellcolor[gray]{0.8} 37MB \\
        & scale = 10 & \fcellsec{2.06} & \fcellsec{1.27} & \fcellsec{2.01}  & \cellcolor[gray]{0.9} 10.8KB & \cellcolor[gray]{0.9} 108KB & \cellcolor[gray]{0.8} 41MB \\
          & scale = 100 & \fcellsec{2.24}  & \fcellsec{1.63} & \fcellsec{2.21} & \cellcolor[gray]{0.9} 12.4KB & \cellcolor[gray]{0.9} 124KB & \cellcolor[gray]{0.8} 47MB  \\
          & scale = 1000 &  \fcellsec{1.09}& \fcellsec{0.7} & \fcellsec{1.07} & \cellcolor[gray]{0.9} 5.2KB& \cellcolor[gray]{0.9} 52KB & \cellcolor[gray]{0.8} 19.8MB\\
         \hline 
         \hline
         \multirow{6}{*}{\textbf{QuickShift}} & $\sigma=0, m=10$  & \fcellsec{44.8} & \fcellsec{28.4} & \fcellsec{48} & \cellcolor[gray]{0.9} 270KB & \cellcolor[gray]{0.8} 2.7MB & \cellcolor[gray]{0.5} 1GB \\
          & $\sigma=0, m=20$  & \fcellsec{13.9} & \fcellsec{8.86} & \fcellsec{13.8} & \cellcolor[gray]{0.9} 69KB & \cellcolor[gray]{0.9} 690KB &\cellcolor[gray]{0.8}  263MB \\
          & $\sigma=1, m=10$ & \fcellsec{30.5} & \fcellsec{18.7} & \fcellsec{30.4} & \cellcolor[gray]{0.9} 173KB & \cellcolor[gray]{0.8} 1.7MB & \cellcolor[gray]{0.8} 659MB   \\
          & $\sigma=1, m=20$ &  \fcellsec{7.77} & \fcellsec{5.41}  & \fcellsec{7.29} & \cellcolor[gray]{0.9} 32KB & \cellcolor[gray]{0.9} 317KB &  \cellcolor[gray]{0.8} 120MB\\
         & $\sigma=2, m=10$ & \fcellsec{21.3} & \fcellsec{13.2} & \fcellsec{20.9} & \cellcolor[gray]{0.9} 117KB  & \cellcolor[gray]{0.8} 1.2MB  & \cellcolor[gray]{0.8}  446MB \\
         & $\sigma=2, m=20$ & \fcellsec{4.3} & \fcellsec{3.46} & \fcellsec{4.3}  & \cellcolor[gray]{0.9} 13KB & \cellcolor[gray]{0.9} 128KB & \cellcolor[gray]{0.8}  49MB\\
         \hline
         \hline
          \multirow{4}{*}{\textbf{SLIC}} & n\_segments = 32 &  \fcellsec{2.17} & \fcellsec{1.42} & \fcellsec{3.24} &  \cellcolor[gray]{0.9} 11KB & \cellcolor[gray]{0.9} 113KB & \cellcolor[gray]{0.8}  43MB \\
        & n\_segments = 64 & \fcellsec{5} & \fcellsec{3.26} & \fcellsec{5.14} & \cellcolor[gray]{0.9} 28KB & \cellcolor[gray]{0.9} 287KB & \cellcolor[gray]{0.8} 109MB \\
          & n\_segments = 128 & \fcellsec{9.75} & \fcellsec{5.96}  & \fcellsec{9.98} &\cellcolor[gray]{0.9} 58KB & \cellcolor[gray]{0.9} 579KB&  \cellcolor[gray]{0.8}  220MB\\
         \hline 
         \hline
         \multicolumn{2}{|c|}{\textbf{ML-LOO}} & \cellcolor[gray]{0.4} \texttildelow 1min20s &  \texttildelow \fcellsec{45}  &\cellcolor[gray]{0.4} \texttildelow 1min22s &  \cellcolor[gray]{0.9} \texttildelow 524 KB & \cellcolor[gray]{0.8}  \texttildelow 5.24 MB & \cellcolor[gray]{0.4}  \texttildelow 2GB  \\
         \hline 
    \end{tabular}
    
\end{table*}



\section{Conclusion}
\label{concl}
ML-LOO approach is a simple defense to wipe out a majority of adversarial samples that was extensively tested against  different attack techniques and mixtures of parameters.   Nevertheless, the computational cost and memory requirements cannot be disregarded. In this paper, we proposed a resource optimization approach based on segmentation. Our approach permits a considerable improvement in resource efficiency while maintaining the original main property of ML-LOO: a simple and efficient adversarial detection that quickly wipes out most adversarial samples. In future work, we aim at applying our approach and experiment on other datasets, image datasets with higher resolution in particular. We also want to experiment with more adversarial techniques and study the transferability properties of our approach.
\section*{Acknowledgements}
This work was supported partially by a grant from the university research board of the American  university of Beirut (URB-AUB-2020/2021). The authors would like to thank Joshua Peter Handali, Johannes Schneider and Pavel Laskov from University of Liechtenstein for the early discussion of the paper topic.

\bibliographystyle{ACM-Reference-Format}
\bibliography{main}
\end{document}